\documentclass[11pt]{article}

\usepackage{amssymb,bbm}
\usepackage{amsfonts}
\usepackage{latexsym}
\usepackage{amsmath}
\usepackage{amsthm}
\usepackage{geometry} 
\usepackage{graphicx}
\usepackage{verbatim}

\usepackage{graphicx}
\usepackage{amsmath,amsthm,amsfonts,amssymb, nccmath}
\usepackage[lofdepth,lotdepth]{subfig}
\newtheorem*{definition*}{Definition}

\usepackage{varwidth}
\usepackage{tabularx,booktabs}
\newcolumntype{C}{>{\centering\arraybackslash}X} 
\usepackage{color}
\usepackage{comment}
\usepackage{gensymb}
\usepackage{bigints}
\usepackage[linesnumbered,lined,boxed,commentsnumbered, ruled]{algorithm2e}
\graphicspath{{./CCST/}{./CST/}}
\def\lddots{\mathinner{\mkern1mu\raise1pt\hbox{.}\mkern2mu  
\raise4pt\hbox{.}\mkern2mu\raise7pt\vbox{\kern7pt\hbox{.}}\mkern1mu}}
\makeatletter
\def\numberbysection{\@addtoreset{equation}{section}
 \def\theequation{\thesection.\arabic{equation}}}
\makeatother  
  
\numberbysection

\newcommand{\be}{\begin{eqnarray}}  
\newcommand{\ee}{\end{eqnarray}}

 \title{\bf On the design of a CST system and its extension to a bi-imaging modality}
\author{ \textsf{C\'ecilia Tarpau$^{a,b,c}$,}
\textsf{Javier Cebeiro$^d$,}
\textsf{Mai K. Nguyen$^a$,}
\textsf{Genevi\`eve Rollet$^b$}
\textsf{and Laurent Dumas $^c$}
\\
\textit{\footnotesize $^{a}$ETIS (CNRS UMR 8051), Universit\'e de Cergy-Pontoise, F-95302 Cergy-Pontoise, France} 
\\
\textit{\footnotesize $^{b}$ LPTM (CNRS UMR 8089),Universit\'e de Cergy-Pontoise, F-95302 Cergy-Pontoise, France} 
\\
\textit{\footnotesize $^{c}$ LMV (CNRS UMR 8100),Universit\'e de Versailles Saint-Quentin, F-78035 Versailles, France}
\\ 
\textit{\footnotesize $^{d}$ CEDEMA, Universidad Nacional de San Mart\'in, B-1650 San Mart\'in, Argentina} \\}
\date{}

\clearpage
\begin{document}

\maketitle
\thispagestyle{empty}
\abstract{{\footnotesize In material testing applications, Computed Tomography is a well established imaging technique that allows the recovery of the attenuation map of an object. Conventional modalities exploit only primary radiation and although in the energy ranges used in industrial applications Compton scatter radiation is significant, it is removed from the measured data. On the contrary, Compton Scattering Tomography not only accounts for the Compton effect but also uses it to image material electronic density.
Despite its promising applications, some aspects of Compton scattering tomography have not been studied so far and, since they are necessary for the design of a operational system, they need to be addressed. 

In this paper we analyze the effect of some physical influences regarding real detectors for the Circular Compton Scattering tomography, a system recently introduced by the authors.  
This is accomplished through the formulation of a new weighted Radon transform. In addition, we propose to adapt the acquired data with pre-processing steps so that they are suitable for reconstruction with a filtered back-projection type reconstruction algorithm, and thus, properly deal with missing data of real measurements. Finally, we introduce a bi-imaging configuration that allows recovering simultaneously the electronic density map of the object as well as its attenuation map.
} }
\section{Introduction}

\begin{figure*}[!ht]
      \begin{center}

    \subfloat[]{
      \centering
    \includegraphics[scale = 0.45]{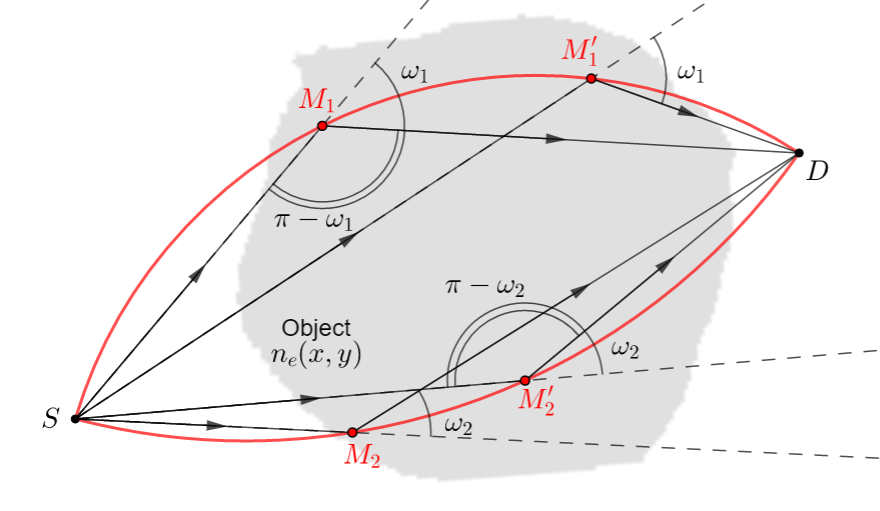}
    \label{fig:cst_concept}
                         } 
    \end{center}
    
    \caption{Concept of CST.}
  \label{fig:ccst}
\end{figure*}

Computed Tomography has been widely used in non-destructive testing and evaluation in industry. This imaging technique exploits the reduction of the intensity of transmitted rays due to the attenuation in matter when X or $\gamma$-rays penetrate deeply on it along linear paths. 
However, CT, as well as other conventional emission imaging techniques, uses only transmitted radiation information. The idea to use also Compton scattered radiation, the predominant photon interaction in the range from 0.1 up to 5 MeV, came up years ago and gave birth to the concept of Compton Scattering Tomography (CST) \cite{lale1959examination, clarke1969compton, farmer1971new}. The image formation process exploited in CST systems rests on the formula linking the scattering angle $\omega$ to the energy  $E_\omega$ of a scattered photon according to
\begin{equation}
    E_\omega = \frac{E_0}{1+k(1-\cos{\omega})},
    \label{eq:formule_compton}
\end{equation}

\noindent
where$E_0$ is the energy of the primary photon, $k$ stands for $E_0/mc^2$ and $m_ec^2=511keV$ is the energy of an electron at rest. This one-to-one correspondence ensures that the considered photon was scattered on a site located on a circular arc whose end-points are the source $S$ and the detector $D$, see Fig. \ref{fig:cst_concept}. A higher electron density results in a higher probability for a photon meeting an electron and consequently a higher scattering count. The idea of using Compton scattering to explore hidden structures of the object, i.e. its electron density, arises as a natural extension of this observation.

CST opens new perspectives for tomographic imaging in industry and exhibits advantages in some scenarios like the scanning of large objects \cite{norton94,evans1998nondestructive, cebeiro_ipse_improved} with source and detectors on the same side. Other applications are airport baggage control \cite{webber2020compton}, cultural heritage inspection \cite{harding2010compton_patrimoine, guerrero2018modellingheritage} and biomedical imaging \cite{redler2018compton_lung, jones2018characterization, nguyen10_trac2}. Each design has its own setup namely moving or fixed source, shape of the detector array, collimators, etc., and its own implementation advantages. In this context, a new two-dimensional CST system has been proposed by the authors \cite{tarpau19trpms, tarpau2020compton}. This two-dimensional modality is made of a fixed source and a ring of detectors passing through the source (see Fig. \ref{fig:ccst_setup_param}). It was called Circular Compton Scattering Tomography (CCST). Preliminary studies on this system reveal some attractive features relative to previously proposed systems. Since CCST does not require any rotation or translation movement of the complete setup in order to obtain a full set of projection data, a reduction of acquisition time in front of other designs can be expected in practice. Furthermore, mechanical complexity is reduced by using non-moving components. In addition, compared to some linear configurations, a circular detector array allows having a compact setup capable of scanning small as well as large objects \cite{tarpau2020compton}. 

So far, the literature on CST has focused mainly on the mathematical aspects of Radon operators modelling it such as inverse formulas \cite{norton94, cebeiro_ipse_improved,nguyen10_trac2, livre_cst, rigaud2012novel_ipse, truong2011radon, webber2018three}, injectivity \cite{Ambartsoumian_2005, agranovsky1996injectivity} or range conditions \cite{finch2006range, ambartsoumian2006range, agranovsky2007range}. Some physical issues of practical importance have been let aside in models namely optimal source positioning, size and energy resolution of detectors, multiple scattering, etc. The permanent progress on detector technology \cite{patentGE, patentSiemens} highlights the need of an accurate characterization of the effects of real detectors on the data and also on reconstructions. 

In previous works on CCST, we also used a theoretical approach of the modality with usual drastic assumptions such as ideal source and detectors and absence of attenuation in matter, as is usual in this field. These studies lead to an exact reconstruction algorithm. Numerical experiments allowed us to prove that information of scattered radiation recovered using CCST is enough to reconstruct the original object. In this article we continue to study the CCST from a practical and more realistic point of view. Particularly, we put spotlight on the effects that real features of detectors like limited energy resolution and finite size, have on reconstruction quality. The introduction of such real factors in the model introduces uncertainty for the localization of the scattering site and missing data. We propose in this article a pre-correction data technique in order to address the problem of missing data.   Futhermore, attenuation is also included in the model.  
With this contribution, we expect boosting up the development of an operating CST imaging system.

The paper is organized as follows. Section \ref{sec:setup} introduces the parametrization of the CCST setup as well as the formulation of data acquisition. Section \ref{sec:presentation_bi-imaging} is devoted to the presentation of a novel bi-imaging system, which combines CCST with a traditional modality of CT.  Numerical experiments as well as strategies to deal with missing data are shown in Section \ref{sec:numerical_experiments}. 
Finally, a discussion about open questions is proposed in Section \ref{sec:open_questions} and concluding remarks of Section \ref{sec:conclusion} end the paper.   
The details of the demonstration of the forward model are given in Appendix \ref{sec:annexe}.

\section{A non-moving system of Compton scattering tomography}
\label{sec:setup}

In this section we explain a configuration for Compton scattering tomography that we have recently introduced in \cite{tarpau19trpms}. Here, we propose a new model for it that takes into account the effect of real detectors on the recorded data. The modality has been called Circular Compton Scatter Tomography and has attractive features like using fixed source and detectors in order to reduce the mechanical complexity of the design. 

\subsection{A fixed setup}
\begin{figure}[!ht]
    \centering
    \includegraphics[scale = 0.55]{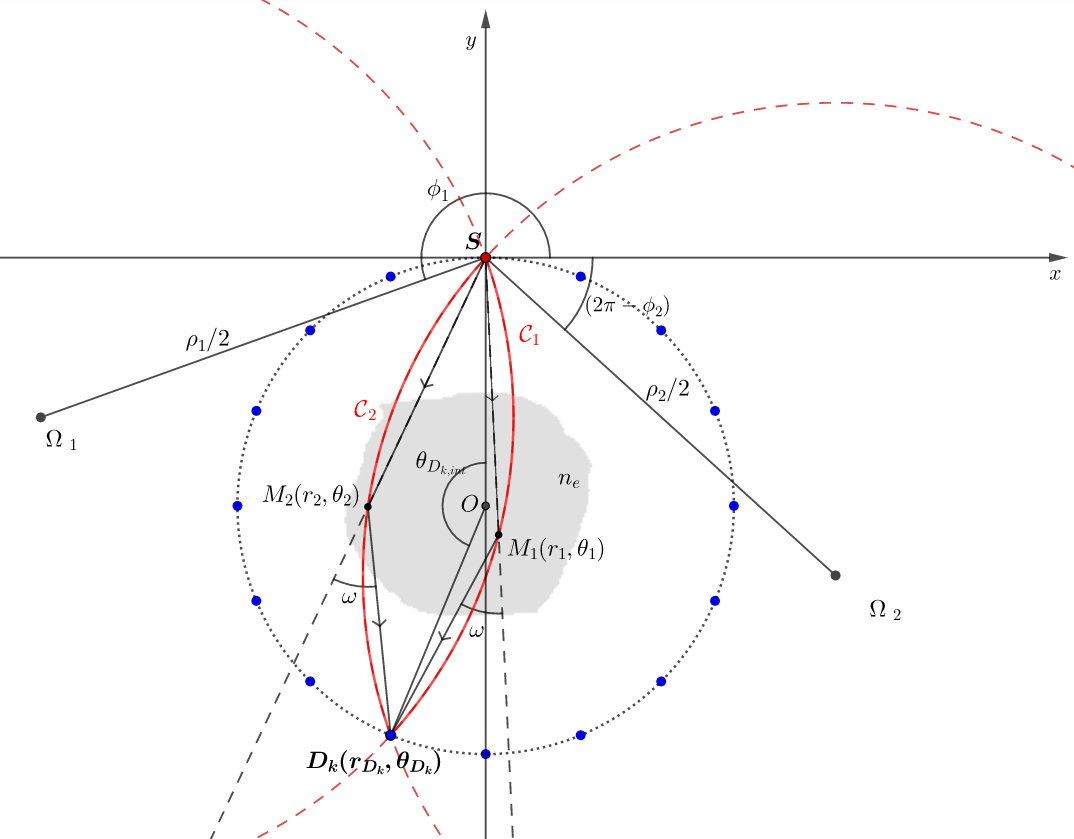}
    \caption{Parametrization of the CCST setup and the scanning arcs of circle}
    \label{fig:ccst_setup_param}
\end{figure}

The system employs a monochromatic source that irradiates an object placed inside of a ring of detectors, see Fig. \ref{fig:ccst_setup_param}. The source $S$ is located at the origin and collimated in order to scan a cross section of the object. Detector cells $D_k$ are distributed along the ring. When a detector registers a scattered photon of energy $E_\omega$ that has been emitted by the source with energy $E_0$, the interaction sites are located on a circle arc passing through $S$ and $D_k$. Then, the flux of photons with energy $E_\omega$ registered by $D$ is proportional to a weighted integral of the electronic density of the object along a pair of circle arcs labelled by the scattering angle $\omega$, see Fig. \ref{fig:ccst_setup_param}. The weighting factors are related to physical effects studied in this work.

\noindent   
Let's consider the ring of diameter $P$ defined by the polar equation $r = P\cos(\theta+\pi/2)$, where $\theta \in [\pi, 2\pi[$. The ring contains $N_D$ detectors $D_k$ labelled by $k\in\{1,.. ,N_D\}$. The position of the center of each detector is given by coordinates $(r_{D_k}, \theta_{D_k})$. In the proposed modelling,  a constant distance between neighbouring detectors is considered. This point is achieved with the definition of $\theta_{D_{k,int}}= 2\pi k/(N_D+1)$, the angle subtended by the $y$-axis and $OD_k$, and related with polar angle $\theta_{D_k}$ with the expression $\theta_{D_{k}} = \pi + \theta_{D_{k,int}}/2$.
Hence, polar coordinates of a detector $D_k$ are 
\begin{equation}
    (r_{D_k} , \theta_{D_k}) = \left( P\cos\left(\theta_{D_k}+\frac{\pi}{2}\right), \pi\left(1+ \frac{k}{N_D+1}\right)\right).
\end{equation}

\noindent 
Detectors are characterized by the arc length $L$ encompassed by them. Finite-size detectors are then modelled by extending its angular limits on both sides of $\theta_{D_{k,int}}$ with $\Delta D$
\begin{equation}
    \theta_{D_{k,int},r} \in [\theta_{D_{k,int}}-\Delta D, \theta_{D_{k,int}}+\Delta D].
\end{equation}
There is a maximum value for $\Delta D$ to avoid \textit{superposition} of detectors in the design of the setup $\Delta D\leq \Delta D_{max} = \pi/(N_D+1), $ which corresponds to the condition on the arc length $ L \leq L_{max} = 2P\Delta D_{max}$.
 
As it was indicated, when a detector $D_k$ of the ring collects a photon with energy $E_\omega$, all the possible scattering sites may be on two circular arcs $\mathcal{C}_1$ and $\mathcal{C}_2$ of equations 
\begin{equation}
    \mathcal{C}_i(\rho_i(\omega,\theta_{D_k}), \phi_i(\omega,\theta_{D_k})): r_i = \rho_i(\omega,\theta_{D_k})\cos{(\theta_i-\phi_i(\omega,\theta_{D_k}))}, \; \; \; i\in \{1,2\}
\end{equation}
where 
\begin{equation}
    \rho_1(\omega, \theta_{D_k})= P\cos(\theta_{D_k}+\pi/2)/\sin(\omega)=\rho_2(-\omega, \theta_{D_k}), \quad \phi_1(\omega,\theta_{D_k}) = \theta_{D_k}+\omega-\pi/2 = \phi_2(-\omega,\theta_{D_k}),
\end{equation}
$\theta_1\in[\theta_{D_k}, \theta_{D_k} +\omega]$ and $\theta_2\in[\theta_{D_k}-\omega, \theta_{D_k}]$. See Fig. \ref{fig:ccst_setup_param}.  For the sake of readability of next equations, we will write $\rho_i$ and $\phi_i$ in reference to $\rho_i(\omega, \theta_{D_k})$ and $\phi_i(\omega,\theta_{D_k})$.
 As we are going to see in next section, circle arcs $\mathcal{C}_1$ and $\mathcal{C}_2$ are the manifold of the transform modeling the modality.

\subsection{A \textit{weighted} Radon transform modelling data}
\label{sec:data_acquisition}

We denote $I$ the intensity of the flux of photons with energy $E_\omega$ collected by a detector $D_k$. From now and then, we denote $I_i$ the intensity of scattered photons of order $i$. The quantity $I$ can be expanded according to the scattering order of the photons, $I=I_0 + I_1+ I_2 + ...$. The model of the CST exploits first-order scattering, the other scattering orders, less predominant than first order, are considered as noise. The flux $I_1$ due to the interaction of photons with electrons in the cross-section of an object with electronic density $n_e$ admits for general expression
\begin{equation}
    I_1(D_k, E_\omega) = \int_{M\in(\mathcal{C}_1\cup\mathcal{C}_2)} a_1(SM, E_0)\,n_e(M)\,q(M, D_k, \omega)\,a_2(MD_k, E_\omega)\,dl(M),
\end{equation}
\noindent 
where $D_k$ is a given detector and $E_\omega$ the registered energy. The weighting function $q$ groups several physical quantities such as solid angle and differential Compton cross section per solid angle. Attenuation factors $a_1$ and $a_2$ correspond to the emitted and transmitted rays along the respective linear paths $SM$ and $MD$, 
\begin{equation}
a_1(SM,E_0)=\exp{\left(-\int_{(x,y)\in SM} \mu_{E_0}(x,y)dl\right)} \quad \text{and} \quad a_2(MD_k,E_{\omega})=\exp{\left(-\int_{(x,y) \in MD_k} \mu_{E_{\omega}}(x,y)dl\right)}.
\end{equation} 
 
We now introduce a formulation for $I_1$ suitable for numerical calculation. First, attenuation coefficients along the linear paths $SM$ and $MD$ are written as negative exponential functions
\begin{equation}
    a_1(E_0; r, \theta ) = \exp{\left(-\int_0^1 \mu_{E_0}\left (t\,x_M, t\,y_M\right)dt\right)} 
\end{equation}
\begin{equation}
    a_2(\theta_{D_k}, E_\omega;r, \theta) = \exp{\left(-\int_0^1 \mu_{E_\omega}\left(x_M + t\left(x_D-x_M\right), y_M+t\left( y_D-y_M\right)\right)dt\right)},
\end{equation}
where $(x_M,y_M) = (r\cos\theta,r\sin\theta)$ and $(x_D,y_D) = (-P\sin{(2\theta_{D_k})}/2 , -P\sin{\theta_{D_k}}^2) $ refer to the respective Cartesian coordinates of the interaction site $M$ and the detector $D$. The weighting function $q$ for the CCST modality  is the same for both scanning circle arcs $\mathcal{C}_1(\rho_1, \phi_1)$ and $\mathcal{C}_2(\rho_2, \phi_2)$ and admits for expression 
\begin{equation}
    q(\theta_{D_k}, \omega;r, \theta) = s(\theta)\frac{d\sigma_c}{d\Omega}(E_0, \omega)\frac{E_0}{E_\omega^2\,k\sin{\omega}}\frac{P|\sin{\theta_{D_k}}\sin(\theta-\theta_{D_k})|}{\sin{\omega}^2}\frac{\sin{(\theta-\phi_1(\theta_{D_k}, \omega))}}{\rho_1(\theta_{D_k}, \omega)^2\cos{(\theta-\theta_{D_k})}^2} \Delta E,
\end{equation}
with $s(\theta)$ is the angular distribution of emitted photons and $\Delta E$ the energetic resolution of the detectors, the reader can refer to Appendix A for the detailed derivation of this factor. Finally, we use Dirac distribution to restrict the integration to the corresponding manifold and $I_1$ is written as a function of $(\theta_{D_k}, \omega)$ according
\begin{multline}
    I_1(\theta_{D_k}, \omega) =\\ \sum_{i=1,2}(-1)^i\int_{\theta_{D_k}}^{\theta_{D_k}+(-1)^i\omega}\,\int_0^\infty 
    a_1(E_0; r, \theta)\,n_e(r, \theta)\, q(\theta_{D_k}, \omega; r, \theta)\, a_2( \theta_{D_k},E_\omega; r, \theta)\,\delta(r-\rho_i\cos(\theta-\phi_i)) \,dr d\theta.
     \label{eq:acq_CCST}
\end{multline}

The latter expression of $I_1$ is a \textit{weigthed} Radon transform on a family of circular arcs. This family considers circular arcs having a fixed extremity at the origin of the coordinates system and the other is located on a circular arc passing through the origin. Equation
 \eqref{eq:acq_CCST} models the data recorded by detectors in this configuration taking into account several physical effects. Consequently, an exact recovery of the map $n_e(r,\theta)$ requires an inversion formula of this weighted Radon transform.

\section{A novel bi-imaging system combining CST and fan-beam CT}
\label{sec:presentation_bi-imaging}

The circular geometry of our configuration allows to associate CCST and conventional fan-beam CT in a single imaging system.
Thus, when detectors are set for recovering transmitted photons of energy $E_0$, the system is used as the well-known fan-beam CT scanner\footnote{The fan-beam CT mode requires, of course, the mechanical rotation of the system.}. Hence, with the dual configuration, one can recover either cross-sections of attenuation map or the electronic density with the CST mode.  With such a bi-imaging system, the majority of the information of recovered photons, that is $I_0$ and $I_1$, is exploited.


\begin{figure*}[!ht]
    \begin{center}
          \subfloat[Fan-beam CT mode]{
      \centering
    \includegraphics[scale = 0.4]{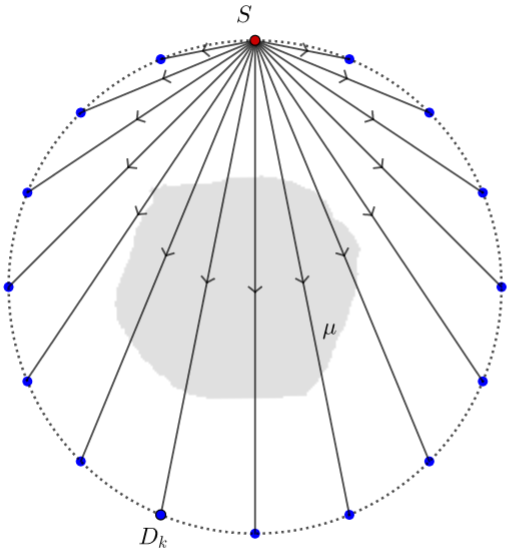}\hfill
    \label{fig:CCST_point}
                          }\hspace{3cm}
    \subfloat[CST mode]{
      \centering
    \includegraphics[scale = 0.4]{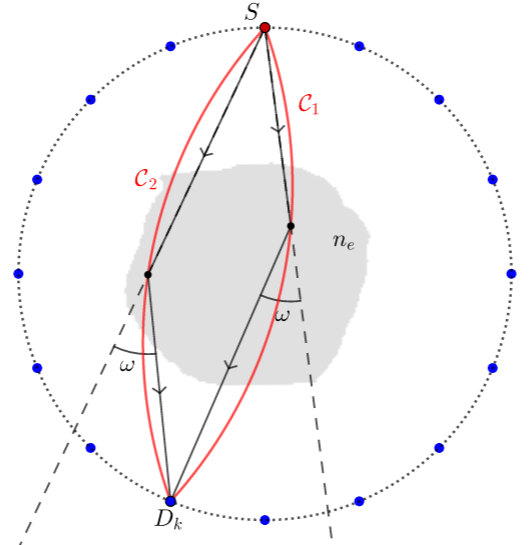}
    \label{fig:ccst_setup}
                         }
    \end{center}

    \caption{The two scanning modes of the bi-imaging system : (a) Fan-beam CT mode and (b) CST mode}
  \label{fig:CCST}
\end{figure*}

\section{Numerical experiments}
\label{sec:numerical_experiments}

We present in this section numerical simulations of the CST mode according to the proposed model of Section \ref{sec:setup}. These numerical experiments focus on CCST, because fan-beam CT has been deeply studied in the last century and is now well known. 

After the presentation of the proposed reconstruction algorithm, some pre-correction steps are proposed to deal with the occurred missing data and reduce its impact on reconstruction quality.


\subsection{Simulated phantom and general parameter choices for the setup}
A phantom of $10\times 10$ cm made of carbon and aluminum was employed for simulation. The object consists of eight circles (see Fig. \ref{fig:setup&obj}) whose parameters are given in Table \ref{table:param_obj}. The geometry of this phantom, albeit simple, allows the visual evaluation of useful criteria such as contrast, spatial resolution (with the circles of different sizes) or the consequences of missing data on the reconstruction of singularities of the object in any direction. 
The electronic densities of carbon and aluminium are set respectively equal to $1.00\cdot10^{23}$ and $6.02\cdot10^{22}$ electrons per cm$^3$. To avoid large numbers of order $10^{23}$ on measurement data, we used preferably the relative electronic density value $\eta$ defined as 
\begin{equation}
    \eta = \frac{n_e}{n_{e, water}}, 
\end{equation}
where $n_{e, water} = 3.23 \cdot 10^{23}$ electrons per cm$^3$ denotes the electronic density of water. Furthermore, attenuation coefficient values of carbon and aluminum between $E_0$ and $E_\pi$ were computed by linear interpolation of ground-truth data of NIST database \cite{nist_database}. 


\begin{table}[!ht]
\small
  \begin{varwidth}[b]{0.55\linewidth}
  \begin{tabular}{r c c c c}
  \toprule
    No. of circle & \multicolumn{2}{c}{Coordinates of the center} & Radius  & Component \\
     & $x$ & $y$ & (cm) &  \\
  \midrule
    1 & 0 & -10 & 5 & Al\\
    2 & 0 & -10 & 4 & C\\
    3 & 2.5 & -10 & 1.1 & Al\\
    4 & 1.25 & -7.8 & 1 & Al\\
    5 & -1.25 & -7.8 & 0.89 & Al\\
    6 & -2.5 & -10 & 0.77 & Al\\
    7 & -1.25 & -12.2 & 0.63 & Al\\
    8 & 1.25 & -12.2 & 0.45 & Al\\
  \bottomrule
  \end{tabular}
   \caption{\small Parameters of the phantom.}
    \label{table:param_obj}
  \end{varwidth}%
  \hfill
  \begin{minipage}[b]{0.4\linewidth}
    \centering
    \includegraphics[scale=0.25]{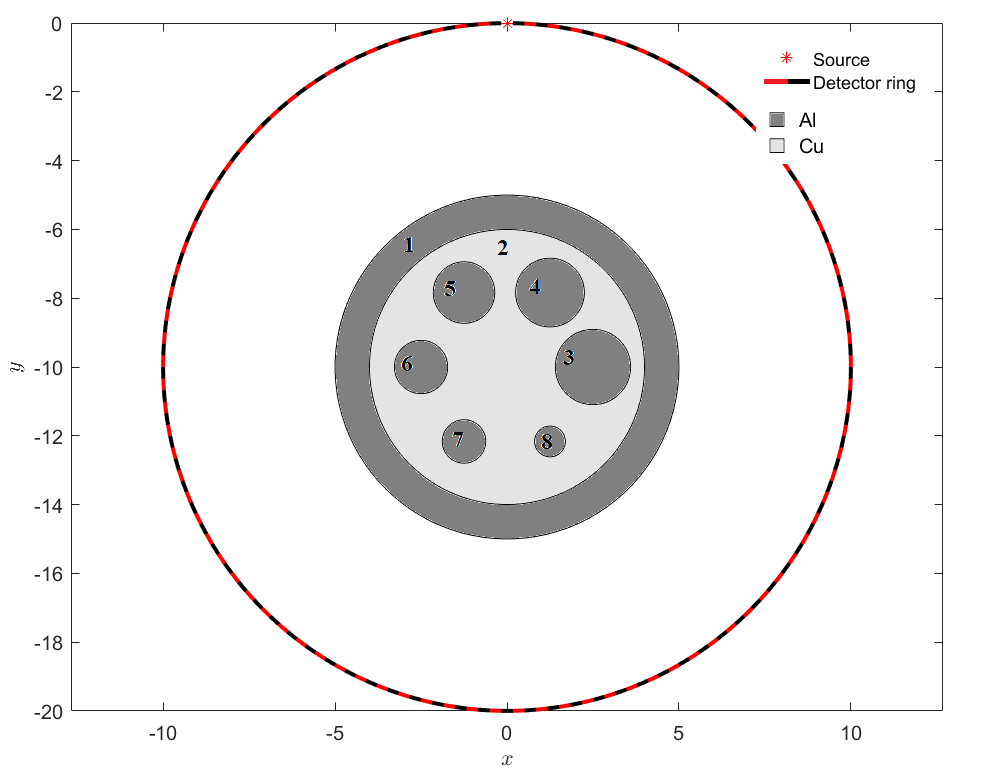}
    \captionof{figure}{\small Setup of the system and simulation phantom}
    \label{fig:setup&obj}
  \end{minipage}
\end{table}

Regarding the simulated setup, we assumed an isotropic and mono-energetic point-like source emits photons of energy $E_0 = 300$ keV. This source was considered as ideal to highlight consequences of real detectors. 
The object was placed inside the ring of diameter $P = 20$ cm containing $N_D = 185$ detectors. This amount represents three detectors per centimeter on the ring. The energy resolution of detectors is set to $\Delta E = 1$keV.

The parameters, described above, remain fixed for all proposed simulations. 

\subsection{Simulation of data measurement from the proposed Radon transform modelling}

From Eq. \eqref{eq:acq_CCST}, some factors of the weighting function can be absorbed into $I_1$ or $n_e$, since they are only functions of $(r, \theta)$ or $(\theta_{D_k}, \omega)$. This avoids also weighting data with very small values (of order $10^{-10}$) of Klein-Nishina differential cross-sections (see Eq. \eqref{eq:klein_nishina}). The weighting function is decomposed into three parts as
\begin{equation}
    q(\theta_{D_k}, \omega\,; r, \theta)  = q_1(\theta_{D_k}, \omega)\, q_2(r, \theta)\, \frac{|\sin{(\theta-\theta_{D_k})}|\sin{(\theta-\phi_1})}{\cos{(\theta-\theta_{D_k})}^2},
\end{equation}
where $q_1(\theta_{D_k}, \omega)$ and $q_2(r, \theta)$ group respectively all coefficients of $q$ depending on $(\theta_{D_k}, \omega)$ and $(r, \theta)$. 
Introducing the $\overline{I_1} = I_1/q_1$, data recovered by detectors are
\begin{multline}
    \overline{I_1}(\theta_{D_k}, \omega) =\\ \sum_{i=1,2}(-1)^i\int_{\theta_{D_k}}^{\theta_{D_k}+(-1)^i\omega}\,\int_0^\infty 
    a_1(E_0; r, \theta)\,\overline{\eta}(r, \theta)\, \overline{q}(\theta_{D_k}, \omega; r, \theta)\, a_2(\theta_{D_k},E_\omega; r, \theta)\,\delta(r-\rho_i\cos(\theta-\phi_i) \,dr d\theta
     \label{eq:acq_CCST_barre}
\end{multline}
where $\overline{\eta} = n_e\cdot q_2/n_{e, water}$ and $\overline{q} = q/(q_1\cdot q_2)$. Then, Eq. \eqref{eq:acq_CCST_barre} is rewritten considering a relative electronic density in Cartesian coordinates to be closer to a real system. This conversion leads to the following parametrisation for scanning circle arcs
\begin{equation}
    \mathcal{C}_i(\theta_{D_k}, \omega) : (x_i(\gamma), y_i(\gamma)) = \frac{\rho_i}{2}(\cos{\phi_i}+\cos{\gamma}, \sin{\phi_i}+\sin{\gamma}), \; \; \gamma \in \left[\theta_{D_k}-\omega+\frac{3\pi}{2}, \theta_{D_k}+\omega+\frac{3\pi}{2}\right]. 
    \label{eq:carac_cart_C1_C2}
\end{equation}

\noindent Using Eq. \ref{eq:carac_cart_C1_C2} in Eq. \ref{eq:acq_CCST_barre} gives finally the equation of acquired data used for simulation

\begin{multline}
    \overline{I_1}(\theta_{D_k}, \omega) = \sum_{i=1,2} \int_{\theta_{D_k}-\omega + 3\pi/2}^{\theta_{D_k}+\omega-3\pi/2} 
    a_1(E_0; x_i(\gamma), y_i(\gamma))\,\overline{\eta}(x_i(\gamma), y_i(\gamma))\, \overline{q}(\theta_{D_k}, \omega; r, \theta) \,a_2(\theta_{D_k},E_\omega; x_i(\gamma), y_i(\gamma))\,d\gamma
     \label{eq:acq_CCST_barre_cart}
\end{multline}

\begin{figure}[!ht]
      \begin{center}
     \subfloat[Image formation $\overline{I_1}(\theta_{D_k}, E_\omega)$ ]{
      \centering
    \includegraphics[scale = 0.28]{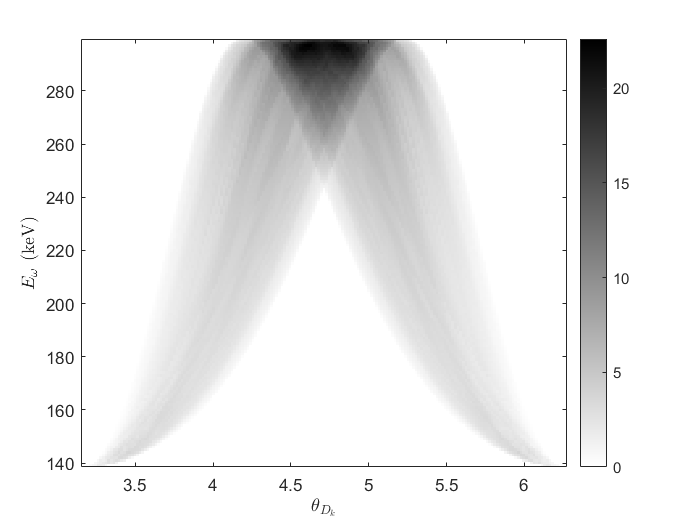}
    \label{fig:acq_C1_C2}
                         }  \hfill
    \subfloat[Image formation $\overline{I_1}(\theta_{D_k}, E_\omega)$ on $\mathcal{C}_1$]{
      \centering
    \includegraphics[scale = 0.28]{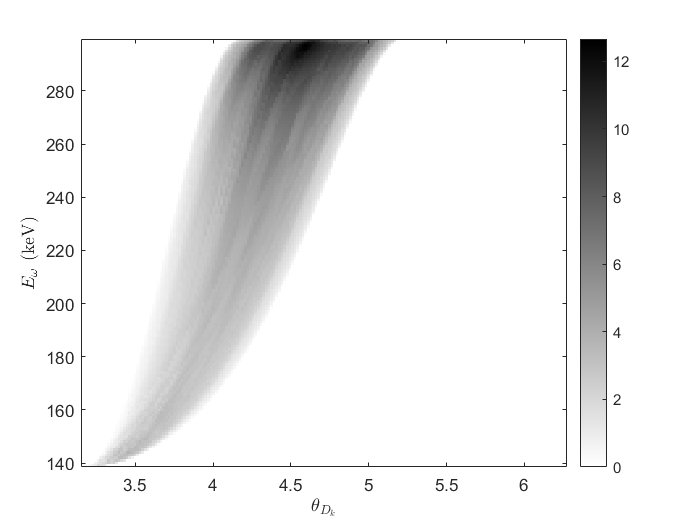}
    \label{fig:acq_C1}
                         }   \hfill
    \subfloat[Image formation $\overline{I_1}(\theta_{D_k}, E_\omega)$ on $\mathcal{C}_2$]{
      \centering
    \includegraphics[scale = 0.28]{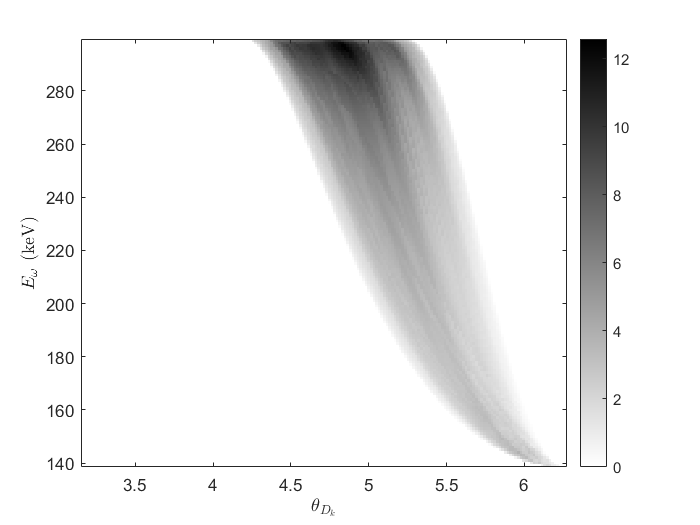}
    \label{fig:acq_C2}
                         }\\
    \subfloat[Rearranged acquired data $\overline{I_1}(\rho, \phi)$]{
      \centering
    \includegraphics[scale = 0.28]{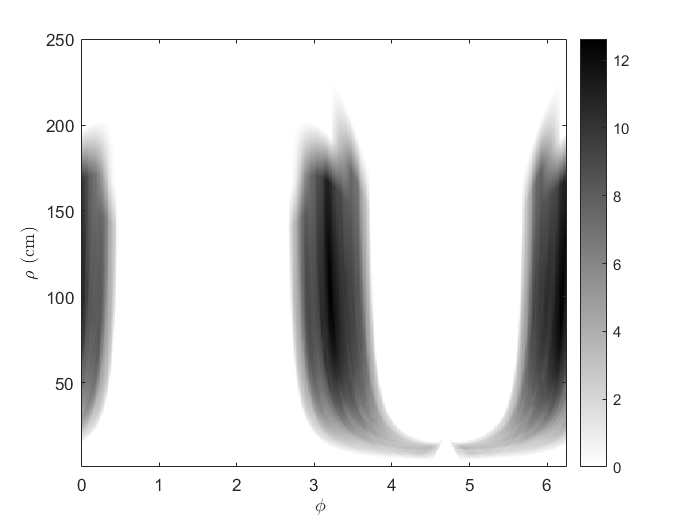}
    \label{fig:acq_rearr}
                         }\hspace{2cm}
    \subfloat[Acquired data under ideal conditions $\widetilde{I_1}(\rho, \phi)$]{
      \centering
    \includegraphics[scale = 0.28]{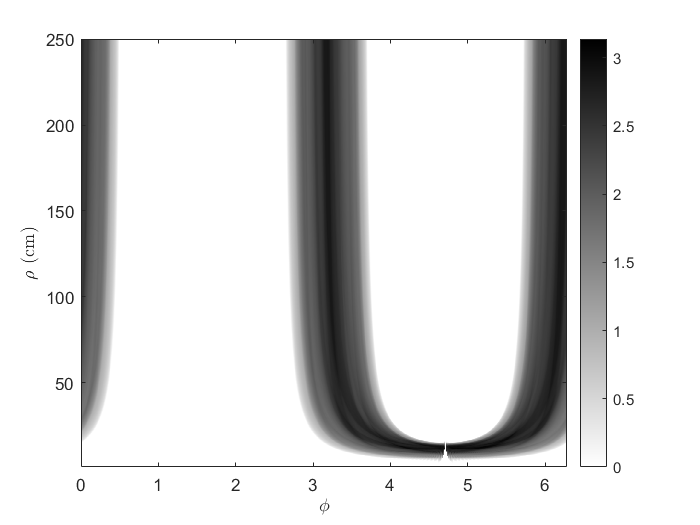}
    \label{fig:acq_ideal}
                         }
    \end{center}
    
    \caption{\small Image acquisition (a) with point-like detectors of energy resolution $\Delta E = 1keV$. Data are separated on $\mathcal{C}_1$ (b) and $\mathcal{C}_2$ (c) and then rearranged (d) for image reconstruction. Comparison with data (e) acquired with \textit{ideal} conditions.} 
\end{figure}

Figure \ref{fig:acq_C1_C2} shows the results of acquired data from Eq. \eqref{eq:acq_CCST_barre_cart} with point-like detectors.


\subsection{On image reconstruction}
\label{sec:image_reconstruction}

The analytic inversion of Eq. \eqref{eq:acq_CCST}, as well as the analytic inversions of the Radon transforms modelling data recorded by other CST systems are still an open challenging question. The complexity of this type of inverse problem lies in the fact that these non-linear integral transform, calls for integration on two different manifolds, on circular arcs first (for the scattering model) and on linear paths (to take into account attenuation of matter). 
This explains why, for the sake of mathematical tractability, the modelling of data acquisition has been studied first by the authors under ideal conditions, that is, ideal source and detectors, perfect energy resolution, absence of noise and attenuation \cite{tarpau19trpms, tarpau2020compton}. From Eq. \eqref{eq:acq_CCST}, these assumptions are translated by the drastic and unrealistic simplification 
\begin{equation}
    a_1(r, \theta, E_0) q(r, \theta; \theta_{D_k}, \omega) a_2(r, \theta; \theta_{D_k}, E_\omega) \approx 1.  
\end{equation}

Furthermore, the intrinsic geometry of CCST introduces an ambiguity on the localization of the scattering site, which can be either on $\mathcal{C}_1$ or $\mathcal{C}_2$. In order to overpass both problems, iterative reconstruction methods such as Kaczmarz's algorithm can be considered but at the cost of computation time and memory requirements. 
In this article, we propose to adapt acquired data to be convenient for a FBP-type reconstruction algorithm, previously proposed by the authors in \cite{tarpau19trpms}. Such an algorithm allows preserving a low computation time without any prior information on the object for reconstruction. Regarding the possibility for the photon of being scattered on two circle arcs, we suppose in the proposed simulations that the detector is able to dissociate photons incoming from one or the other circular arc with, for instance, a system of collimation or filtering. This assumption is discussed in Section \ref{sec:open_questions} of the paper. From Fig. \ref{fig:acq_C1_C2}, one can now obtain data on scanning arcs  $\mathcal{C}_1$ or $\mathcal{C}_2$ separately. See the results on Figs. \ref{fig:acq_C1} and \ref{fig:acq_C2} with the simulation parameters described previously. 

This reconstruction algorithm supposes the recorded data are functions of $(\rho_i, \phi_i)$. Figure \ref{fig:acq_rearr} shows the results of rearranged data from both Figs. \ref{fig:acq_C1} and \ref{fig:acq_C2}. With such change of variables, the circle arc families $\mathcal{C}_1$ and $\mathcal{C}_2$ are grouped into one named $\mathcal{C}$ and its parameters are denoted $(\rho, \phi)$ in the rest of the section. For comparison purposes, Fig. \ref{fig:acq_ideal} depicts measurement data with the same setup upon \textit{ideal} conditions. Consequences of realistic energy resolution are two-fold. First, there is an abrupt cut-off of data when $\rho$ increases (equivalently, when energy values are close to $E_0$) whereas theoretical circular arcs having an \textit{infinite} diameter give also information on the object. We observe in a second step the base of the $U$-shape of Figs. \ref{fig:acq_rearr} and \ref{fig:acq_ideal}. In addition to few missing data caused by the distance between detectors adjacent to the source in Fig. \ref{fig:acq_ideal}, fixed energy resolution is responsible for an enlargement of this hole in Fig. \ref{fig:acq_rearr}. Some pre-processing steps can be considered in order to attenuate the consequences of these missing data. Before presenting it, we explain in next paragraph the reconstruction algorithm. 

\subsubsection{Algorithm}

Denoting $\widetilde{I}_1$ the data measurement under ideal conditions, the cross-section of the electronic density $n_e(x,y)$ of an object in Cartesian coordinates can be recovered exactly from $\widetilde{I}_1$ \cite{tarpau19trpms} with 
\begin{equation}
    n_e(x,y) = \frac{1}{2\pi}\;\int_0^{2\pi}d\phi\; \frac{1}{x\cos{\phi}+y\sin{\phi}}\cdot\mathcal{F}^{-1}\left(-i\cdot\text{sign}(\nu)\mathcal{F}\left(\frac{\partial \widetilde{I}_1(\rho, \phi)}{\partial\rho}\rho\right)(\nu)\right)\left(\frac{x^2+y^2}{x\cos{\phi}+y\sin{\phi}}\right), 
    \label{eq:recons_cart_final}
\end{equation}
where $\mathcal{F}$ denotes the one-dimensional Fourier transform. 

Algorithm \ref{algo:reconssss} summarizes the different steps for reconstructing the object from Eq. \eqref{eq:recons_cart_final}.

\IncMargin{1em}
    \begin{algorithm}[!ht]
        \SetAlgoLined
        \KwData{$\widetilde{I}_1(\rho, \phi)$, data measurement of function $n_e(x,y)$ in \textit{ideal} conditions}
        \KwResult{$n_e(x,y)$}
       Compute discrete derivation of $\widetilde{I}_1(\rho, \phi)$ relative to variable $\rho$ and multiply the result by $\rho$\;
        Filter the result in Fourier domain by $-i\cdot\text{sign}(\nu)$ \; 
        For each $\varphi$, interpolate the data on the considered scanning circles of equation $(x^2+y^2)/(x\cos\varphi+y\sin\varphi)=\text{constant}$\;
        Weight the result using the factor $1/(x\cos\varphi+y\sin\varphi)$\;
        Sum the weighted interpolations on all directions $\varphi$\;
        Weight the result by $1/2\pi$\;
        
    \caption{Reconstruction of the object}
    \label{algo:reconssss}
    \end{algorithm}
    \DecMargin{1em}

\subsubsection{Optimal object positioning}
\begin{figure}[!ht]
    \subfloat[ ]{
      \centering
    \includegraphics[scale = 0.3]{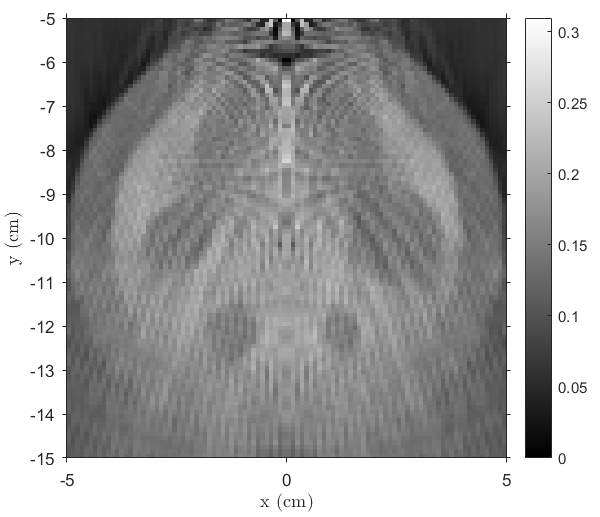}
    \label{fig:milieu}
                         }
    \subfloat[ ]{
      \centering
    \includegraphics[scale = 0.3]{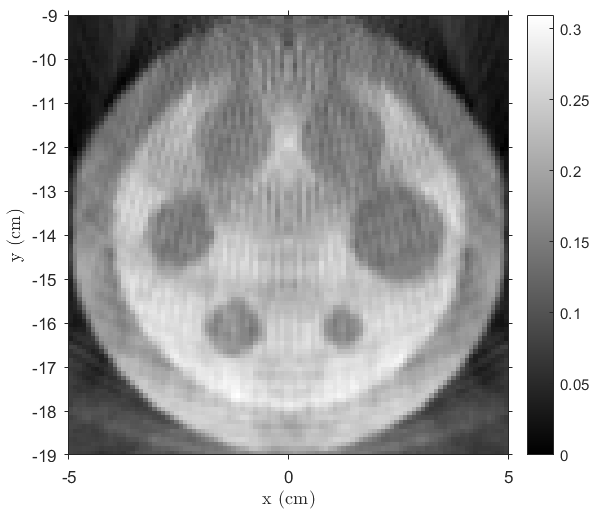}
    \label{fig:pres}
                         }
  \centering
    \subfloat[ ]{
      \centering
    \includegraphics[scale = 0.3]{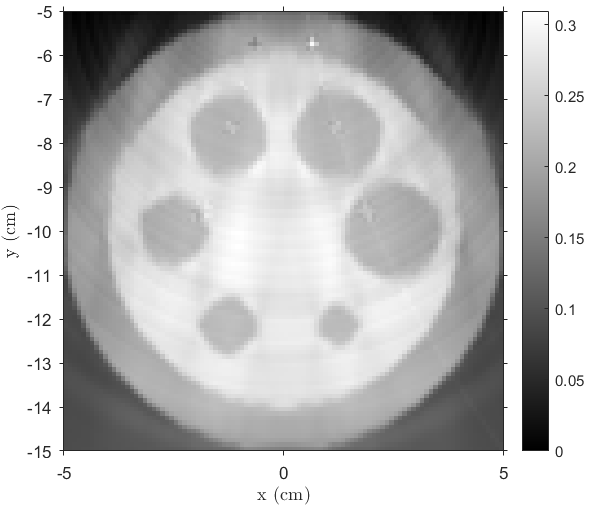}
    \label{fig:objet_th}
                         }
  \caption{\small Optimal object positioning for fixed energy resolution. (a) Reconstruction with object centered on the ring. (b) Reconstruction with object close to detectors opposite from the source.}
  \label{fig:position_obj}
\end{figure}

The proposed algorithm gives its best reconstruction results when the discretisation of variables $(\rho, \phi)$ is uniform \cite{tarpau19trpms}. As an example, Fig. \ref{fig:objet_th} shows the reconstruction result obtained from rearranged data (Fig. \ref{fig:acq_ideal}) ideal conditions. 

However, when the modelling of data acquisition is performed from parameters $(\theta_{D_k}, \omega)$ with a finite and constant energy resolution, missing data is expected since the relation between $\omega$ and $\rho$ is non-linear. Figure \ref{fig:milieu} shows the result of image reconstruction from $\overline{I_1}(\rho, \phi)$ of Fig. \ref{fig:acq_rearr}. The reconstructed image exhibits strong artifacts at its centre and interior circles tend to have rhomboidal shapes. The choice of the optimal position for the object inside the ring is the first way to improve reconstruction quality. In CCST case, this optimal position consists in shifting the object close to detectors opposite from the source. In fact, this operation  allows concentrating data as much as possible on small values of $\omega$, hence with approximately the same discretisation. Figure \ref{fig:pres} shows the results considering the object as its new position.  The object is now well reconstructed as the whole although circle streak artifacts persist on reconstruction. 


\subsubsection{Additional data pre-processing steps for correcting consequences of fixed energy resolution}

\begin{figure}[!ht]
\centering
    \subfloat[ ]{
      \centering
    \includegraphics[scale = 0.3]{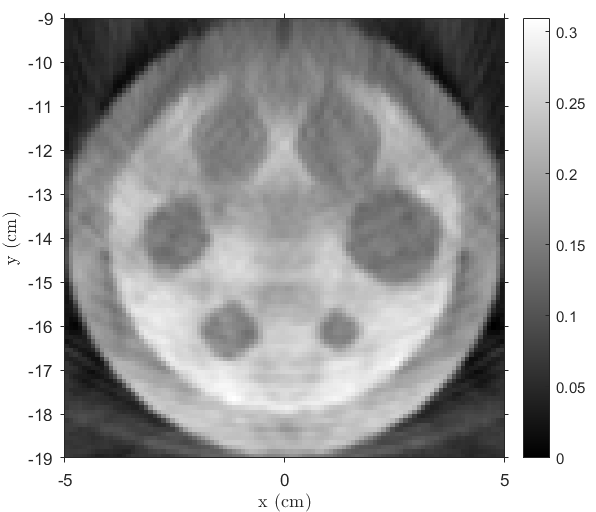}
    \label{fig:filt}
                         }
    \subfloat[ ]{
      \centering
    \includegraphics[scale = 0.3]{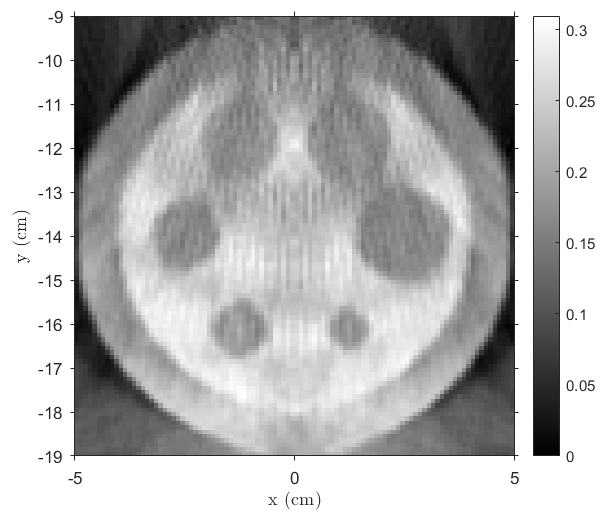}
    \label{fig:interp}
                        }
    \subfloat[ ]{
      \centering
    \includegraphics[scale = 0.3]{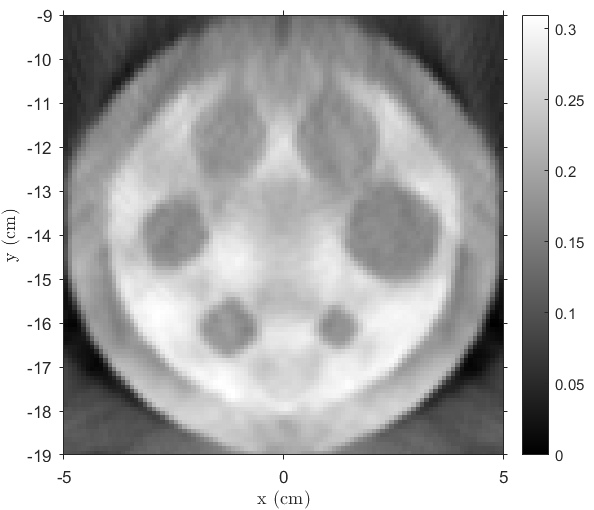}
    \label{fig:filt_interp}
                         }
    \caption{\small Reconstruction after some data pre-processing. (a) Reconstruction after smoothing. (b) Reconstruction after interpolation of missing data. (c) Reconstruction after both pre-processing steps.}
  \label{fig:pre_correction_data}
\end{figure}

To deal with the described artifacts of Fig. \ref{fig:pres}, some pre-processing steps for data can be added. This is first due to abrupt cut-off on data may be diminished by the use of a smoothing filter as shows reconstruction of Fig. \ref{fig:filt}. Second, ring artifacts are reduced via interpolation of missing data. The reconstruction result is depicted in Fig. \ref{fig:interp}. Finally, reconstruction from data after both pre-correction steps is given in Fig. \ref{fig:filt_interp}. These two pre-correction steps allow reducing the majority of artifacts. However, upper interior circles have still rhomboidal shapes. This point can be solved with a finer energy resolution for detectors.

\subsubsection{Influence of finite-size detectors on reconstruction quality}
\begin{figure}[!ht]
\centering
\subfloat[ ]{
      \centering
    \includegraphics[scale = 0.35]{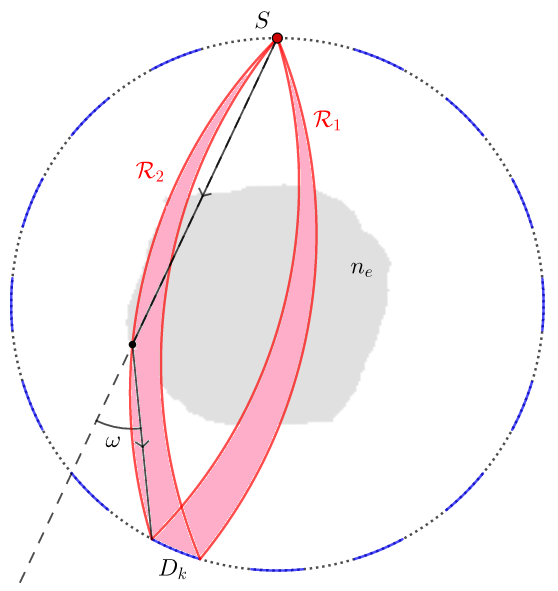}
    \label{fig:ccst_real}
                         }
\subfloat[ ]{
      \centering
    \includegraphics[scale = 0.3]{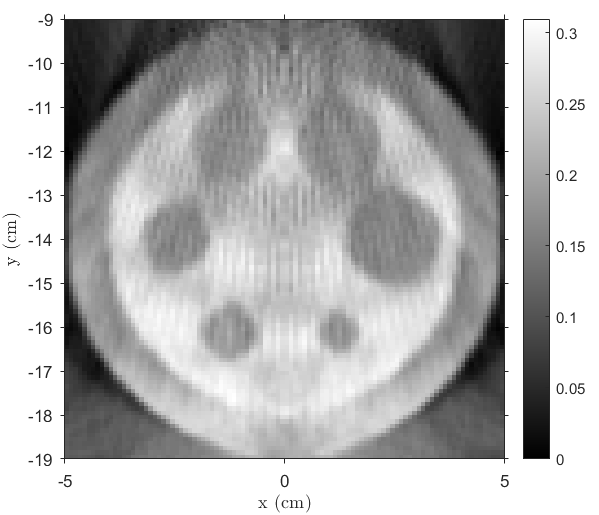}
    \label{fig:lmax_sur_2}
                         }
    \subfloat[ ]{
      \centering
    \includegraphics[scale = 0.3]{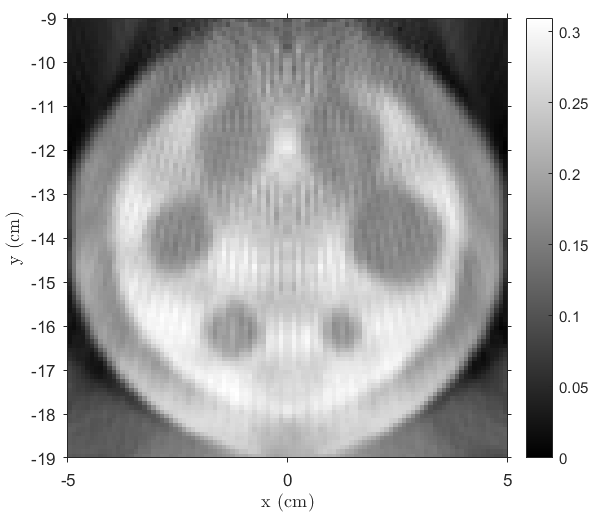}
    \label{fig:lmax}
                         }
        \caption{\small Influence of finite-size detectors: (a) Data acquisition is now made on two regions, namely $\mathcal{R}_1$ and $\mathcal{R}_2$. Reconstruction results for detectors of arc length (b) $L = L_{max}/2$  and (c) $L= L_{max}$ }
\end{figure}

We study now the consequences of finite-size detectors on reconstruction quality. Data acquisitions resulting to the simulations of this paragraph are not pre-processed, in order to not interfere with consequences of such detectors. With this configuration, all the possible scattering sites may be now on two circular regions $\mathcal{R}_1$ and $\mathcal{R}_2$, see Fig. \ref{fig:ccst_real}. The ambiguity on the localization of the scattering site implies blur on reconstructions, see the result for detector lengths $L = L_{max}/2$ and  $L = L_{max}$ on Figs. \ref{fig:lmax_sur_2} and \ref{fig:lmax} and compare them to Fig. \ref{fig:pres}, where we had point-like detectors. 


\section{Discussions}
\label{sec:open_questions}


This study opens some new interesting issues that will be considered in future works. 

\subsection{Practical additional considerations for the model}
The proposed study gives a step towards a practical model for CCST. However, additional features have to be added to complete our simulation setup. 

First, future work will consider real characteristics for the source. Blur, resulting in uncertainty on the localization of scattering sites, is expected to affect reconstruction. As with the issue of finite-size detectors, some deblurring strategies are considered as solutions by the authors for future work. Such methods have already proved their worth previously in photo-acoustic tomography \cite{roitner2014deblurring}.
Furthermore, consequences on reconstruction quality of some additional realistic features of energy sensitive detectors will be studied. As an example, incoming photons to detectors may also be affected by an additional loss of energy due to a Compton effect inside the detector. Even if current and future advances on detector technology are expected to reduce the probability of this kind of event, this phenomenon has to be taken into account. Second, a careful analysis of the collimation on detectors which dissociates photons from a scattering site on $\mathcal{C}_1$ or $\mathcal{C}_2$ has to be considered. Blur is expected if the collimation system is not perfect. Work is also on the way to consider a reconstruction formula able to process data directly from a family of double circular arc to completely remove this collimation. 

Moreover, multiple scattering is for the moment considered as noise in our model. 
A work \cite{rigaud20193d} proposed recently a step further in this direction, with a model which integrates data acquired from second-order scattered data $I_2$. 

Some work is on the way to validate the proposed model performing Monte-Carlo simulations. 

\subsection{Improving reconstruction quality}
This work showed that the FBP-type algorithm which is exact analytically under ideal conditions, is also able to reconstruct despite attenuation and some realistic aspects of detectors. However, there is still progress to be made in terms of the quality of reconstruction. If one wants to use such FBP-type algorithm, additional pre and post-processing steps can be considered. As an example, the iterative attenuation correction algorithm on reconstruction proposed recently by the authors in \cite{tarpau2019ndt} can be added. This method takes advantage of transmitted data $I_0$ of fan-beam CT mode. Secondly, as seen in the previous paragraph, deblurring methods may correct consequences of finite-size sensors and detectors on reconstructions. However, combining many of these corrective treatments may have a counter-productive effect on reconstruction quality. In that case, iterative reconstruction techniques or recently introduced deep learning approaches may offer a good alternative. Otherwise, an interesting work in \cite{rigaud20193d} proposes a feature reconstruction algorithm that can be adapted to this modality. 

\subsection{Extension of CCST in three dimensions}
Extensions of CCST in three dimensions are also a subject for future work. Two geometries can be considered: one fixed source and detectors on a sphere or on a cylinder passing through the source. These geometries allow preserving the advantages of the two dimensional system. Modelling image acquisition using these 3D geometries leads to Radon transforms on tori, and their inversion are still an open challenging problem. A preliminary study about these modalities have been proposed recently by the authors in \cite{tarpau20203d}.

\section{Concluding remarks}
\label{sec:conclusion}
 
In this paper we revisit a model for Compton scatter tomography in order to incorporate some physical influences that are relevant in a real scenario namely energy resolution of detectors, finite size cells, photoelectric attenuation, etc. In addition, we introduced a bi-imaging system that enables to reconstruct simultaneously the electronic density as well as the attenuation map of the object. Numerical experiments reveal the drawbacks of the reconstruction technique face to missing data and allow to explore new ways to overcome this emerging difficulties like pre-correction of data, optimal object positioning or detector design. The approach is a clear step towards a high fidelity modeling for optimal design before a physical implementation. Monte-Carlo simulations will complement the model by including higher order scattering, some work is on the way.

\appendix
\section{Derivation of the weighted RT for real source and detectors}
\label{sec:annexe}

\subsection{General formulation of $dN$, the number of photons collected by a detector}
\begin{figure}[!ht]
      \begin{center}
         =
      \centering
    \includegraphics[scale = 0.45]{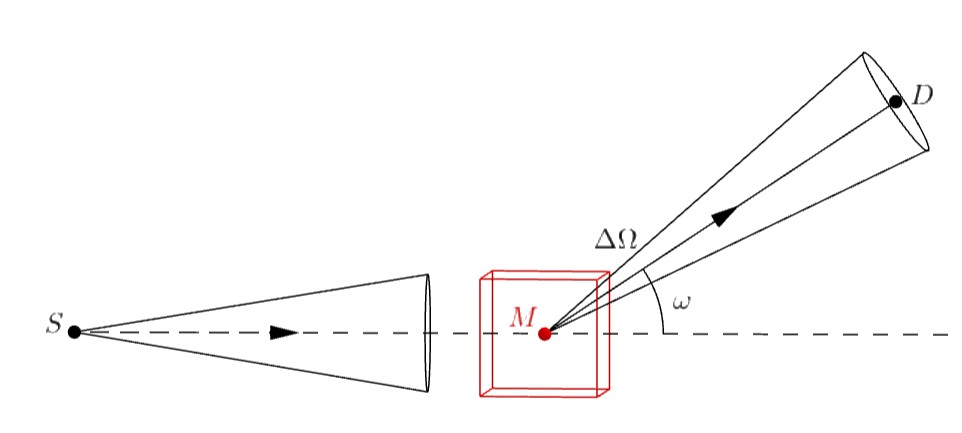}
    \label{fig:C_effect_prin}
    \end{center}
    
    \caption{\small Solid angle in Compton effect.}
  \label{fig:ccst_solid_angle}
\end{figure}

We consider a beam of photons emitted by a source at energy $E_0$ that undergo scattering with angle $\omega$ when they reach an electron. Then, scattered photons continue through its path and some reach a detector $D$. Neglecting attenuation coefficients due to emission and scattering phases of the photon trajectory, the number of photons $dN$ collected by a detector is given by the relation 
\begin{equation}
    dN = F\, d\sigma_c\,n_e dV, 
    \label{eq:initial_dN}
\end{equation}
where $F$ is the incident photon flux, $n_e dV$ is the number of electrons in the small volume $dV$ and $d\sigma_c$ is the differential Compton cross section. The distribution of photons with respect to the scattering angle $\omega$ into a solid angle $d\Omega$ (see Fig. \ref{fig:ccst_solid_angle}) is given by the Klein-Nishina differential cross section
\begin{equation}
    \frac{d\sigma_c}{d\Omega}(E_0, \omega) = \frac{r_0^2}{2}\left(\frac{E_\omega}{E_0}\right)^2\left(\frac{E_\omega}{E_0}+\frac{E_0}{E_\omega}-\sin{\omega}^2\right),
    \label{eq:klein_nishina}
\end{equation}
where $r_0 = hc\alpha/2\pi m_ec^2$ is the classical electron radius with $h$ the Plank constant, $c$ the speed of light, $\alpha$ the fine structure constant, $m_e$ the electron mass and $E_\omega$ the energy of the photon after collision. 

We now introduce in Eq. \eqref{eq:initial_dN} the differential Compton cross-section per solid angle $d\sigma_c/d\Omega$ 
\begin{equation}
    dN = F\, \frac{d\sigma_c}{d\Omega}\,n_e\,d\Omega dV. 
    \label{eq:initial_dN_avec_dOmega}
\end{equation}

\subsection{Derivation of the weighting factor $q(r, \theta; \omega, \theta_D)$}

We scan planar cross-sectional maps of thickness $\Delta z$, hence we can assume that the photon flux and electronic density of the object are invariant along the $z$-direction. The incident photon flux $F$ can be expressed as
\begin{equation}
    F(r,\theta,z) = \frac{s(\theta)}{r\Delta z},
    \label{eq:expF}
\end{equation}
where $s(\theta)$ stands for the angular density of emitted photons by the source. Substituting $dV$ by $rdrd\theta\Delta z$ and according to \eqref{eq:expF}, the number of photons incoming from a scattering site $M$ of polar coordinates $(r, \theta)$ collected by a detector $D(r_D, \theta_D)$ is rewritten as 
\begin{equation}
    dN(r, \theta, \theta_D) = s(\theta) \,\frac{d\sigma_c}{d\Omega}(E_0, \omega(r, \theta, \theta_D))\, n_e(r,\theta)\,d\Omega drd\theta. 
\end{equation}

Given a scattering angle $\omega$, we saw in section \ref{sec:setup} that two scanning arcs $\mathcal{C}_1$ or $\mathcal{C}_2$ may correspond. For this demonstration, we group these two family into one, denoted $\mathcal{C}$, introducing the variable $\beta \in ]-\pi, \pi[\backslash\{0\}$.
\begin{equation}
   \mathcal{C}(\rho(\beta,\theta_D), \phi(\beta,\theta_D)): r_1 = \rho(\beta,\theta_D)\cos{(\theta_1-\phi(\beta,\theta_D))}, \; \; \;  \theta_1\in[\theta_D, \theta_D +\beta] 
\end{equation}
When $\beta$ is positive, it describes the family $\mathcal{C}_1$, otherwise, it refers to  $\mathcal{C}_2$. 

\noindent Then, with the successive changes of variables from $(r, \theta, \theta_D)$ to $(\beta, \theta, \theta_D)$ and then $(E_\beta, \theta, \theta_D)$, we arrive to the following expression of the count rate 
\begin{equation}
    dN(E_\beta,\theta, \theta_d) = S(\theta)\,\frac{d\sigma_c}{d\Omega}(E_0, \beta)\, n_e(r(E_\beta,\theta, \theta_D), \theta)\,\frac{E_0}{E_\beta^2\,k\sin{\beta}}\, \frac{P|\sin{\theta_D}\sin(\theta-\theta_D)|}{\sin{\beta}^2}d\Omega dEd\theta
\end{equation}

Integrating $dN(E_\beta,\theta, \theta_D)$ over $\theta$, one can obtain $I(E_\beta, \theta_D)$, the intensity of photons at energy $E_\beta$ with a step $\Delta E$ collected by a detector $D$ into the solid angle $\Delta\Omega$:
\begin{align}
    I(E_\beta, \theta_D) &= \int_{\theta_D-\beta}^{\theta_D+\beta} dN(E_\beta,\theta, \theta_d) \nonumber\\
    &= \int_{\theta_D-\beta}^{\theta_D+\beta} S(\theta)\,\frac{d\sigma_c}{d\Omega}(E_0, \beta)\, n_e(r(E_\beta,\theta, \theta_D), \theta)\,\frac{E_0}{E_\beta^2\,k\sin{\beta}}\, \frac{P|\sin{\theta_D}\sin(\theta-\theta_D)|}{\sin{\beta}^2}\Delta\Omega \Delta E d\theta.
\end{align}

Introducing the delta function-kernel, one has 

\begin{multline}
    I(E_\beta, \theta_D) = \int_{\theta_D-\beta}^{\theta_D+\beta} S(\theta)\,\int_0^\infty \frac{d\sigma_c}{d\Omega}(E_0, \beta)\, n_e(r, \theta)\,\frac{E_0}{E_\beta^2\,k\sin{\beta}}\\ \frac{P|\sin{\theta_D}\sin(\theta-\theta_D)|}{\sin{\beta}^2} \delta(r-\rho(\beta, \theta_D)\cos(\theta-\phi(\beta, \theta_D)))\Delta\Omega \Delta E dr d\theta.
\end{multline}

Going back to scattering angle $\omega$ and introducing a weighting function $q(r, \theta; \omega, \theta_D) $ as 
\begin{equation}
    q(r, \theta; \omega, \theta_D) = s(\theta)\frac{d\sigma_c}{d\Omega}(E_0, \omega),\frac{E_0}{E_\omega^2\,k\sin{\omega}}\frac{P|\sin{\theta_D}\sin(\theta-\theta_D)|}{\sin{\omega}^2}\Delta\Omega \Delta E
\end{equation}
one gets 
\begin{multline}
    I(\omega, \theta_D) = \int_{\theta_D-\omega}^{\theta_D} \int_0^\infty  n_e(r, \theta)q(r, \theta; -\omega, \theta_D)  \delta(r-\rho( -\omega, \theta_D)\cos(\theta-\phi(-\omega, \theta_D))) dr d\theta \,+ \\
     \int_{\theta_D}^{\theta_D+\omega} \int_0^\infty  n_e(r, \theta)q(r, \theta; \omega, \theta_D)\delta(r-\rho( \omega, \theta_D)\cos(\theta-\phi(\omega, \theta_D))) dr d\theta
\end{multline}


Equation \ref{eq:acq_CCST} is the response obtained from detectors using CCST. Finally, one can obtain the expression of solid angle $\Delta\Omega$ from Fig.\ref{fig:ccst_setup_param}
\begin{equation}
    \Delta\Omega(r, \theta;  \omega, \theta_D) = \frac{a}{4\pi}\frac{\boldsymbol{\overrightarrow{MD_k}\cdot\overrightarrow{n}}}{||\boldsymbol{\overrightarrow{MD_k}}||^3} = \frac{a}{4\pi}\frac{\sin{(\theta-\phi(\omega, \theta_D))}}{\rho(\omega, \theta_D)^2\cos(\theta-\theta_D)^2},
\end{equation}
where $a$ is the detector area and $\boldsymbol{\overrightarrow{n}}$ the unitary normal vector to the considered detector.







\bibliographystyle{IEEEtran}
\bibliography{cas-refs}

\begin{thebibliography}{10}
\providecommand{\url}[1]{#1}
\csname url@samestyle\endcsname
\providecommand{\newblock}{\relax}
\providecommand{\bibinfo}[2]{#2}
\providecommand{\BIBentrySTDinterwordspacing}{\spaceskip=0pt\relax}
\providecommand{\BIBentryALTinterwordstretchfactor}{4}
\providecommand{\BIBentryALTinterwordspacing}{\spaceskip=\fontdimen2\font plus
\BIBentryALTinterwordstretchfactor\fontdimen3\font minus
  \fontdimen4\font\relax}
\providecommand{\BIBforeignlanguage}[2]{{%
\expandafter\ifx\csname l@#1\endcsname\relax
\typeout{** WARNING: IEEEtran.bst: No hyphenation pattern has been}%
\typeout{** loaded for the language `#1'. Using the pattern for}%
\typeout{** the default language instead.}%
\else
\language=\csname l@#1\endcsname
\fi
#2}}
\providecommand{\BIBdecl}{\relax}
\BIBdecl

\bibitem{lale1959examination}
P.~Lale, ``The examination of internal tissues, using gamma-ray scatter with a
  possible extension to megavoltage radiography,'' \emph{Physics in Medicine \&
  Biology}, vol.~4, no.~2, p. 159, 1959.

\bibitem{clarke1969compton}
R.~Clarke and G.~Van~Dyk, ``Compton-scattered gamma rays in diagnostic
  radiography,'' in \emph{Medical Radioisotope Scintigraphy. VI Proceedings of
  a Symposium on Medical Radioisotope Scintigraphy}, 1969.

\bibitem{farmer1971new}
F.~Farmer and M.~P. Collins, ``A new approach to the determination of
  anatomical cross-sections of the body by compton scattering of gamma-rays,''
  \emph{Physics in Medicine \& Biology}, vol.~16, no.~4, p. 577, 1971.

\bibitem{norton94}
S.~J. Norton, ``Compton scattering tomography,'' \emph{Journal of applied
  physics}, vol.~76, no.~4, pp. 2007--2015, 1994, [doi:10.1063/1.357668].

\bibitem{evans1998nondestructive}
B.~L. Evans, J.~Martin, L.~Burggraf, and M.~Roggemann, ``Nondestructive
  inspection using {C}ompton scatter tomography,'' \emph{IEEE Transactions on
  Nuclear Science}, vol.~45, no.~3, pp. 950--956, 1998.

\bibitem{cebeiro_ipse_improved}
J.~Cebeiro, M.~K. Nguyen, M.~Morvidone, and A.~Noumow{\'e}, ``New
  “improved” {C}ompton scatter tomography modality for investigative
  imaging of one-sided large objects,'' \emph{Inverse Problems in Science and
  Engineering}, vol.~25, no.~11, pp. 1676--1696, 2017,
  [doi:10.1080/17415977.2017.1281920].

\bibitem{webber2020compton}
J.~Webber and E.~L. Miller, ``Compton scattering tomography in translational
  geometries,'' \emph{Inverse Problems}, vol.~36, no.~2, p. 025007, 2020.

\bibitem{harding2010compton_patrimoine}
G.~Harding and E.~Harding, ``Compton scatter imaging: {A} tool for historical
  exploration,'' \emph{Applied Radiation and Isotopes}, vol.~68, no.~6, pp.
  993--1005, 2010, [doi:10.1016/j.apradiso.2010.01.035].

\bibitem{guerrero2018modellingheritage}
P.~Guerrero, M.~K. Nguyen, L.~Dumas, and S.~X. Cohen, ``Modelling of a new
  {C}ompton imaging modality for an in-depth characterisation of flat heritage
  objects,'' in \emph{Proceedings of The 9th EUROSIM Congress on Modelling and
  Simulation 2016}, no. 142.\hskip 1em plus 0.5em minus 0.4em\relax
  Link{\"o}ping University Electronic Press, 2018, pp. 334--340,
  [doi:10.3384/ecp17142334].

\bibitem{redler2018compton_lung}
G.~Redler, K.~C. Jones, A.~Templeton, D.~Bernard, J.~Turian, and J.~C. Chu,
  ``Compton scatter imaging: {A} promising modality for image guidance in lung
  stereotactic body radiation therapy,'' \emph{Medical physics}, vol.~45,
  no.~3, pp. 1233--1240, 2018, [doi:10.1002/mp.12755].

\bibitem{jones2018characterization}
K.~C. Jones, G.~Redler, A.~Templeton, D.~Bernard, J.~V. Turian, and J.~C. Chu,
  ``Characterization of {C}ompton-scatter imaging with an analytical simulation
  method,'' \emph{Physics in Medicine \& Biology}, vol.~63, no.~2, p. 025016,
  2018, [doi:10.1088/1361-6560/aaa200].

\bibitem{nguyen10_trac2}
M.~K. Nguyen and T.~T. Truong, ``Inversion of a new circular-arc {R}adon
  transform for {C}ompton scattering tomography,'' \emph{Inverse Problems},
  vol.~26, no.~6, p. 065005, 2010, [doi:10.1088/0266-5611/26/9/099802].

\bibitem{tarpau19trpms}
C.~Tarpau, J.~Cebeiro, M.~Morvidone, and M.~K. Nguyen, ``A new concept of
  {C}ompton {S}cattering tomography and the development of the corresponding
  circular {R}adon transform,'' \emph{IEEE Transactions on Radiation and Plasma
  Medical Sciences}, vol. (accepted for publication), 2019,
  [10.1109/TRPMS.2019.2943555].

\bibitem{tarpau2020compton}
C.~Tarpau and M.~K. Nguyen, ``Compton scattering imaging system with two
  scanning configurations,'' \emph{Journal of Electronic Imaging}, vol.~29,
  no.~1, p. 013005, 2020.

\bibitem{livre_cst}
T.~Truong and M.~K. Nguyen, ``Recent developments on {C}ompton scatter
  tomography: theory and numerical simulations,'' in \emph{Numerical
  Simulation-From Theory to Industry}.\hskip 1em plus 0.5em minus 0.4em\relax
  IntechOpen, 2012, [doi:10.5772/2600].

\bibitem{rigaud2012novel_ipse}
G.~Rigaud, M.~K. Nguyen, and A.~K. Louis, ``Novel numerical inversions of two
  circular-arc {R}adon transforms in {C}ompton scattering tomography,''
  \emph{Inverse Problems in Science and Engineering}, vol.~20, no.~6, pp.
  809--839, 2012, [doi:10.1080/17415977.2011.653008].

\bibitem{truong2011radon}
T.~T. Truong and M.~K. Nguyen, ``Radon transforms on generalized {C}ormack's
  curves and a new {C}ompton scatter tomography modality,'' \emph{Inverse
  Problems}, vol.~27, no.~12, p. 125001, 2011.

\bibitem{webber2018three}
J.~W. Webber and W.~R. Lionheart, ``Three dimensional {C}ompton scattering
  tomography,'' \emph{Inverse Problems}, vol.~34, no.~8, p. 084001, 2018.

\bibitem{Ambartsoumian_2005}
\BIBentryALTinterwordspacing
G.~Ambartsoumian and P.~Kuchment, ``On the injectivity of the circular {R}adon
  transform,'' \emph{Inverse Problems}, vol.~21, no.~2, pp. 473--485, feb 2005.
  [Online]. Available:
  \url{https://doi.org/10.1088$\%$2F0266-5611$\%$2F21$\%$2F2$\%$2F004}
\BIBentrySTDinterwordspacing

\bibitem{agranovsky1996injectivity}
M.~L. Agranovsky and E.~T. Quinto, ``Injectivity sets for the {R}adon transform
  over circles and complete systems of radial functions,'' \emph{Journal of
  Functional Analysis}, vol. 139, no.~2, pp. 383--414, 1996.

\bibitem{finch2006range}
D.~Finch \emph{et~al.}, ``The range of the spherical mean value operator for
  functions supported in a ball,'' \emph{Inverse Problems}, vol.~22, no.~3, p.
  923, 2006.

\bibitem{ambartsoumian2006range}
G.~Ambartsoumian and P.~Kuchment, ``A range description for the planar circular
  {R}adon transform,'' \emph{SIAM journal on mathematical analysis}, vol.~38,
  no.~2, pp. 681--692, 2006.

\bibitem{agranovsky2007range}
M.~Agranovsky, P.~Kuchment, and E.~T. Quinto, ``Range descriptions for the
  spherical mean {R}adon transform,'' \emph{Journal of Functional Analysis},
  vol. 248, no.~2, pp. 344--386, 2007.

\bibitem{patentGE}
Y.~Jin, P.~Edic, X.~Rui, and G.~Fu, ``Energy discriminating photon-counting
  detector and the use thereof,'' Patent WO 2018/129\,243 A1, 2018-07-12.

\bibitem{patentSiemens}
A.~H. Vija and M.~Rodrigues, ``Multi-modal compton and single photon emission
  computed tomography medical imaging system,'' Patent WO 2020/032\,922 A1,
  2020-02-13.

\bibitem{nist_database}
\BIBentryALTinterwordspacing
J.~H. Hubbell and S.~M. Seltzer. (2004) {Attenuation Coefficients and Mass
  Energy-Absorption Coefficients (Version 1.4.)}. [Online]. Available:
  \url{http://physics.nist.gov/xaamdi}
\BIBentrySTDinterwordspacing

\bibitem{roitner2014deblurring}
H.~Roitner, M.~Haltmeier, R.~Nuster, D.~P. O’Leary, T.~Berer, G.~Paltauf,
  H.~Gr{\"u}n, and P.~Burgholzer, ``Deblurring algorithms accounting for the
  finite detector size in photoacoustic tomography,'' \emph{Journal of
  biomedical optics}, vol.~19, no.~5, p. 056011, 2014.

\bibitem{rigaud20193d}
G.~Rigaud, ``3d compton scattering imaging: study of the spectrum and contour
  reconstruction,'' \emph{arXiv preprint arXiv:1908.03066}, 2019.

\bibitem{tarpau2019ndt}
C.~Tarpau, J.~Cebeiro, and M.~K. Nguyen, ``A new bi-imaging {N}{D}{T} system
  for simultaneous recovery of attenuation and electronic density maps,'' in
  \emph{Eleventh Int. Conf. Non Destruct. Test. Aerosp}, 2019.

\bibitem{tarpau20203d}
C.~Tarpau, J.~Cebeiro, M.~K. Nguyen, G.~Rollet, and L.~Dumas, ``On {3}{D}
  imaging systems based on scattered ionizing radiation,'' in
  \emph{Unconventional Optical Imaging II}, vol. 11351.\hskip 1em plus 0.5em
  minus 0.4em\relax International Society for Optics and Photonics, 2020, p.
  1135107.

\end{thebibliography}





\end{document}